\documentclass[12pt]{iopart}
\usepackage{iopams} 
\usepackage{setstack} 
\usepackage{hyperref}  
\usepackage{graphicx}
\usepackage{cite}
\usepackage{url}  
\usepackage{bm}
\usepackage{color}

\begin{document}

\title[BEC due to Diffusivity Edge under Periodic Confinement]{Bose-Einstein-like Condensation due to Diffusivity Edge under Periodic Confinement}

\author{Beno\^it Mahault$^1$ \& Ramin Golestanian$^{1,2}$}

\address{$^1$Max Planck Institute for Dynamics and Self-Organization (MPIDS), 37077 G\"ottingen, Germany}
\address{$^2$Rudolf Peierls Centre for Theoretical Physics, University of Oxford, Oxford OX1 3NP, United Kingdom}
\vspace{10pt}
\begin{indented}
\item[]\today
\end{indented}

\begin{abstract}
A generic class of scalar active matter, characterized at the mean field level by the diffusivity vanishing above some threshold density, was recently introduced [Golestanian R 2019 {\it Phys. Rev. E} {\bf 100} 010601(R)]. In the presence of harmonic confinement, such ‘diffusivity edge’ was shown to lead to condensation in the ground state, with the associated transition exhibiting formal similarities with Bose-Einstein condensation (BEC). In this work, the effect of a diffusivity edge is addressed in a periodic potential in arbitrary dimensions, where the system exhibits coexistence between many condensates. Using a generalized thermodynamic description of the system, it is found that the overall phenomenology of BEC holds even for finite energy barriers separating each neighbouring pair of condensates. Shallow potentials are shown to quantitatively affect the transition, and introduce non-universality in the values of the scaling exponents.
\end{abstract}


\section{Introduction}

Systems in which detailed balance is broken at the microscopic scale are commonly referred to as active matter \cite{Gompper2020,RamaswamyAnRev}.
This definition encompasses various processes, which often result in self-propulsion of the microscopic units. When coupled to other mechanisms, activity generally triggers novel physics as it possibly leads to nontrivial types of emergent self-organization \cite{MarchettiRMP,bechinger2016RMP,RG-LesHouches}.

One of the many fascinating properties of active systems is their ability to phase separate even in the absence of explicit attractive interactions. 
A good example of such a feature is the motility-induced phase separation, which emerges when persistent motion is coupled to local motility inhibition \cite{cates2015MIPS,Henkes:2011,Hagan:2013,Buttinoni:2013,soto+golestanian14}, and can lead to the formation of close-packed ordered structures \cite{Blaschke:2016,Letitia:2018,Gompper:2018}. Dilute systems with short-range velocity alignment also exhibit phase separation at the onset of macroscopic orientational order \cite{solon2015phase,Chate2019AnnRev}. Clustering is, moreover, known to arise when the interaction between active particles is induced by a self-generated scalar field (concentration, temperature, etc.) \cite{Golestanian2012ThermalColloids,taktikos2012PRE,Saha2014PRE,liebchen2015PRL,Varma2018SoftMatt,Agudo2019PRL},
and from hydrodynamics \cite{Zottl2014PRL,Blaschke2016SoftMatter}, 
resulting in effective long-range interactions.

In the absence of long-range orientational order \cite{TONER:2005}, the long-time mean field description of the aforementioned systems is commonly achieved via a conservation law for the density field $\rho$, with generic drift and diffusion contributions. The effective mobility and diffusion coefficients that result from coarse-graining are then generally explicit functions of $\rho$. Recently, a new class of scalar active matter was introduced in which the consequences of the existence of a diffusivity edge at a critical concentration $\rho_c$ (i.e. diffusion vanishes when $\rho \geq \rho_c$) was examined \cite{Golestanian2019BEC}. It was discovered that when confined in a harmonic potential, systems falling into this class undergo a transition formally akin to Bose-Einstein condensation (BEC), thus providing a new non-equilibrium mechanism for the emergence of clustering.

There are many examples in which structure formation in active matter results in a pattern formation that involves the selection of a characteristic length-scale, such that the cluster sizes are limited and do not scale with the system size \cite{taktikos2012PRE,Saha2014PRE,liebchen2015PRL,Zottl2014PRL,Tjhung2018PRX}. Such microscopic confinement can be modeled at the mean field level 
by introducing an effective potential which provides multiple sites for condensation. Moreover, periodic potential landscapes are commonly used to manipulate driven colloidal systems. Such periodic potentials are expected to lead to cluster-lattices. Here, we study the phenomenology that arises from a diffusivity edge in such configurations. 

We consider a sinusoidal egg-crate confinement in arbitrary dimension $d$, and identify two limiting regimes for the system. For deep potentials, the system behaves similarly to the case of a single harmonic trap case treated in Ref. \cite{Golestanian2019BEC}. However, we find that the existence of finite energy barriers between neighbouring condensates quantitatively modifies the transition. A generalized thermodynamic description shows that the overall phenomenology of BEC is always preserved. However, for shallow potentials we observe quantitative differences as compared to the classical BEC description. Most notably, we find that the exponent associated with the scaling of the condensate fraction with respect to an effective temperature is non-universal, and depends on how the diffusion scales with $\rho_c - \rho$.

The rest of the paper is organized as follows. We introduce the model in Sec. \ref{sec:model} and characterize the phenomenology of the condensation transition in Sec. \ref{sec:char}. Section \ref{sec_Thermodynamics_cos} is devoted to the development of the generalized thermodynamics associated with the phenomenology of the system, and Sec. \ref{sec_discussion} concludes the paper. 


\section{Scalar active matter with diffusivity edge}\label{sec:model}

We start by introducing the formalism that will be used throughout the paper. 
In the mean field approach considered here, the particle density field $\rho({\bm r},t)$
obeys the following conservation law
\begin{equation}
\partial_t \rho + \nabla \cdot {\bm J} = 0 \,, \qquad {\bm J} = -M(\rho) \rho \nabla U -D(\rho) \nabla \rho \,,
\label{eq_continuity}
\end{equation}
where $U({\bm r})$ denotes the external confining potential. The dynamics conserves the total number of particles $N = \int \rmd^d {\bm r} \, \rho\left({\bm r},t\right)$ in the accessible $d$-dimensional space at all times.
The mobility $M(\rho)$ and diffusion coefficient $D(\rho)$ are in general density-dependent.
Their ratio in the zero-density limit defines a tuning parameter 
\begin{equation}
k_{\rm B} T_{\rm eff} \equiv \frac{D(\rho \to 0)}{M(\rho \to 0)} \,,
\end{equation}
which gives a measure of the fluctuations at the particle level, and can be assimilated to an effective temperature for the system.
Because this study aims at describing systems that are non-equilibrium in essence, the fluctuation-dissipation theorem (FDT) can be broken for finite densities, namely, 
\begin{equation}
\frac{D(\rho)}{M(\rho)} \neq \frac{D(\rho \to 0)}{M(\rho \to 0)} \,. 
\end{equation}
This feature can be interpreted as collective inhibition or activation caused by the interplay of activity and, for instance, interactions. In particular, for sufficiently large densities we assume the existence of a diffusivity edge in the system, defined as\ $D(\rho )/M(\rho) = 0$ for $\rho \ge \rho_c$. The non-local effects due to hydrodynamic interactions in the presence of broken FDT are neglected in our work \cite{Golestanian:2002}.


\begin{figure}[t!]
	\centering
	\includegraphics[width = 0.7\textwidth]{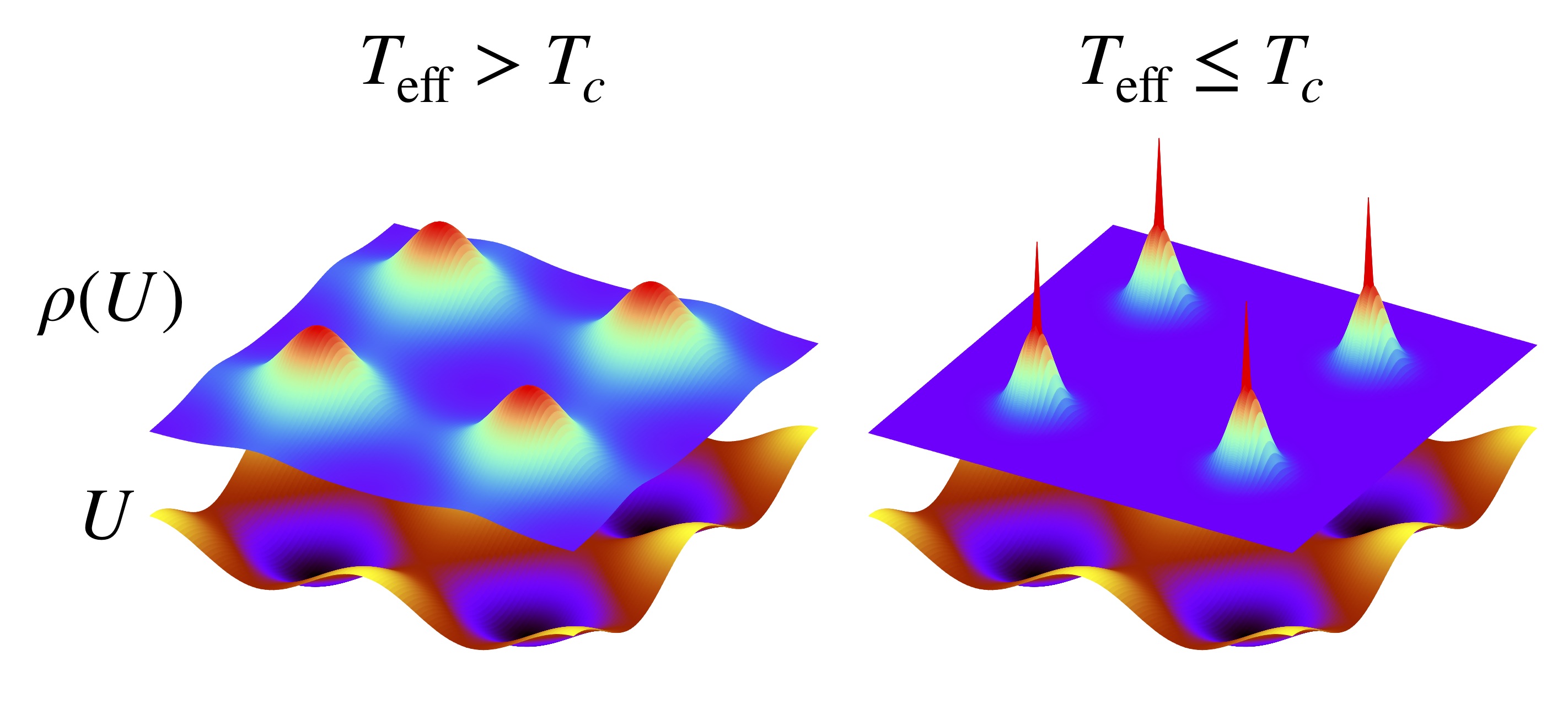}
	\caption{Schematic representation of the solution \eref{eq_Bweights_rho} above and below the 
	condensation transition in a two dimensional periodic potential.
	When the effective temperature is larger than $T_c$, then $\rho_0 < \rho_c$, and $\rho(U)$
	shows a smooth decay from the minimum of the potential.
	Below $T_c$, $\rho(U)$ exhibits a sharp peak at the ground state, which reflects the presence of a condensate.}
	\label{fig_sketch}
\end{figure}


The steady-state solutions of \Eref{eq_continuity} are computed by setting the current to zero (${\bm J}={\bm 0}$), leading to
\begin{equation}
\frac{\rmd U}{\rm d\rho} = - \frac{D(\rho)}{M(\rho)\rho} \,,
\label{eq_solU}
\end{equation}
which can be formally used to obtain $\rho(U)$. The normalization condition in the stationary state can be written as $N = \int \rmd^d {\bm r} \, \rho\left(U({\bm r})\right)=\int \rmd U \, g(U) \rho(U)$, where $g(U)$ is the relevant density of states.

Since $D(\rho)/M(\rho) \ge 0$, we surmise that $\rho$ is a decreasing function of $U$. We denote $\rho_0$ as the maximal value that $\rho$ takes in the ground state $U = 0$. When $\rho_0 < \rho_c$, $\rho(U)$ can be obtained simply by integrating and inverting \Eref{eq_solU}, in which case $\rho_0$ is then determined from the density normalization. When the effective temperature decreases, $\rho_0$ increases until it reaches the maximally allowed value of $\rho_c$. The transition temperature $T_c$ is defined as the value taken by $T_{\rm eff}$ when $\rho_0 = \rho_c$. For $T_{\rm eff} \le T_c$, $\rho$ is thus not a smooth function at $U = 0$ (see Figure \ref{fig_sketch}): we obtain $\rho(U \to 0^+) = \rho_c$ and the value $\rho$ takes in the ground state is undefined, which reflects the formation of a condensate.

In most of this work we will consider for simplicity a step profile for the ratio of diffusion over mobility
\begin{equation}
\frac{D(\rho)}{M(\rho)} = \cases{k_{\rm B} T_{\rm eff} & $\rho < \rho_c$ \\
0 & $\rho \ge \rho_c$} \,.
\end{equation}
Therefore, denoting $\beta \equiv 1/k_{\rm B} T_{\rm eff}$ for convenience, 
the density is given by the Boltzmann weights 
\begin{equation}
\label{eq_Bweights_rho} \rho(U) = 
\cases{\rho_0 \exp(-\beta U) \qquad & $T_{\rm eff} > T_c$ \\ 
N_c\delta(U) g(U)^{-1} + \rho_c \exp(-\beta U) & $T_{\rm eff} \le T_c$ } \,,
\end{equation}
where the contribution $N_c\delta(U) g(U)^{-1}$ ensures the overall normalization in the condensed phase.


\section{Characterization of the condensation transition} \label{sec:char}

\label{sec_Transition}

In this work we consider the sinusoidal potential in $d$ dimensions, as sketched in Figure \ref{fig_sketch} for $d=2$, and defined by
\begin{equation}
U({\bm r}) =  \sum_{k=1}^d \tilde{U}(r_k)\,, \quad \tilde{U}(r_k) = \frac{U_b}{2d}\left[1 - \cos\left(\frac{\pi r_k}{r_b}\right)\right] \,,
\label{eq_cosU}
\end{equation}
where the $r_k$'s denotes the Cartesian coordinates of the position ${\bm r}$. The system is thus divided into identical cells of volume $(2 r_b)^d$, each separated by an energy barrier $U_b$. Assuming an equal partitioning of the particles over the cells, and using the fact that the density can be factorized (as can be seen by combining Equations (\ref{eq_Bweights_rho}) and (\ref{eq_cosU})), we find
\begin{equation}
N  = \cases{ \rho_0 L^d \exp\left(-\frac{\beta U_b}{2}\right) I^d_0\left( \frac{\beta U_b}{2d} \right) & $T_{\rm eff} > T_c$ \\
N_c + \rho_c L^d \exp\left(-\frac{\beta U_b}{2}\right) I^d_0\left( \frac{\beta U_b}{2d} \right) & $T_{\rm eff} \le T_c$ } \,,
\label{eq_N_cos}
\end{equation}
where $L$ denotes the linear system size and $I_\nu(x) = \int_0^\pi \rmd s\, \exp[x \cos(s)]\cos(\nu s)/\pi$ is the modified Bessel function of the first kind of integer rank $\nu$, which has the following asymptotic forms
\begin{equation}
 \label{eq_Bessel}
 \!\!\!\! \!\!\!\!  \!\!\!\! I_\nu(x) \underset{x \to \infty}{\sim} \frac{e^x}{\sqrt{2\pi x}} \left(1 + \Or\left(x^{-1}\right) \right) \,, \qquad
 I_\nu(x)  \underset{x \to 0}{\sim} \frac{1}{\nu !}\left(\frac{x}{2}\right)^{\nu}\left( 1 + \Or\left(x^2\right)\right) \,.
 \end{equation}
Hence, in the strong and weak confinement limits the ground state density below the diffusivity edge obeys
 \begin{equation}
\label{eq_rho0_cos} 
 \rho_0 \underset{k_{\rm B}T_{\rm eff} \ll U_b}{\sim} n \left( \frac{2\pi k_{\rm B}T_{\rm eff}}{k} \right)^{-\frac{d}{2}} \,, \qquad
\rho_0 \underset{k_{\rm B}T_{\rm eff} \gg U_b}{\sim} \frac{N}{L^d} \left( 1 + \Or(\beta U_b) \right) \,,
 \end{equation}
 where $n = N (2 r_b / L)^d$ denotes the number of particles in each cell, 
 and $k =\pi^2  U_b / (2 d r_b^2)$ measures the effective potential stiffness in the ground state.
Because edge effects vanish when the barrier height $U_b$ is much larger than the effective temperature,
the expression given in~\Eref{eq_rho0_cos} for $k_{\rm B}T_{\rm eff} \ll U_b$ is identical to the one derived in Ref.~\cite{Golestanian2019BEC}
for an infinite harmonic trap. On the other hand, when $k_{\rm B}T_{\rm eff} \gg U_b$ the system is dominated by fluctuations and the density reaches a uniform profile.


\begin{figure}[t!]
	\centering
	\includegraphics[width = 0.7\textwidth]{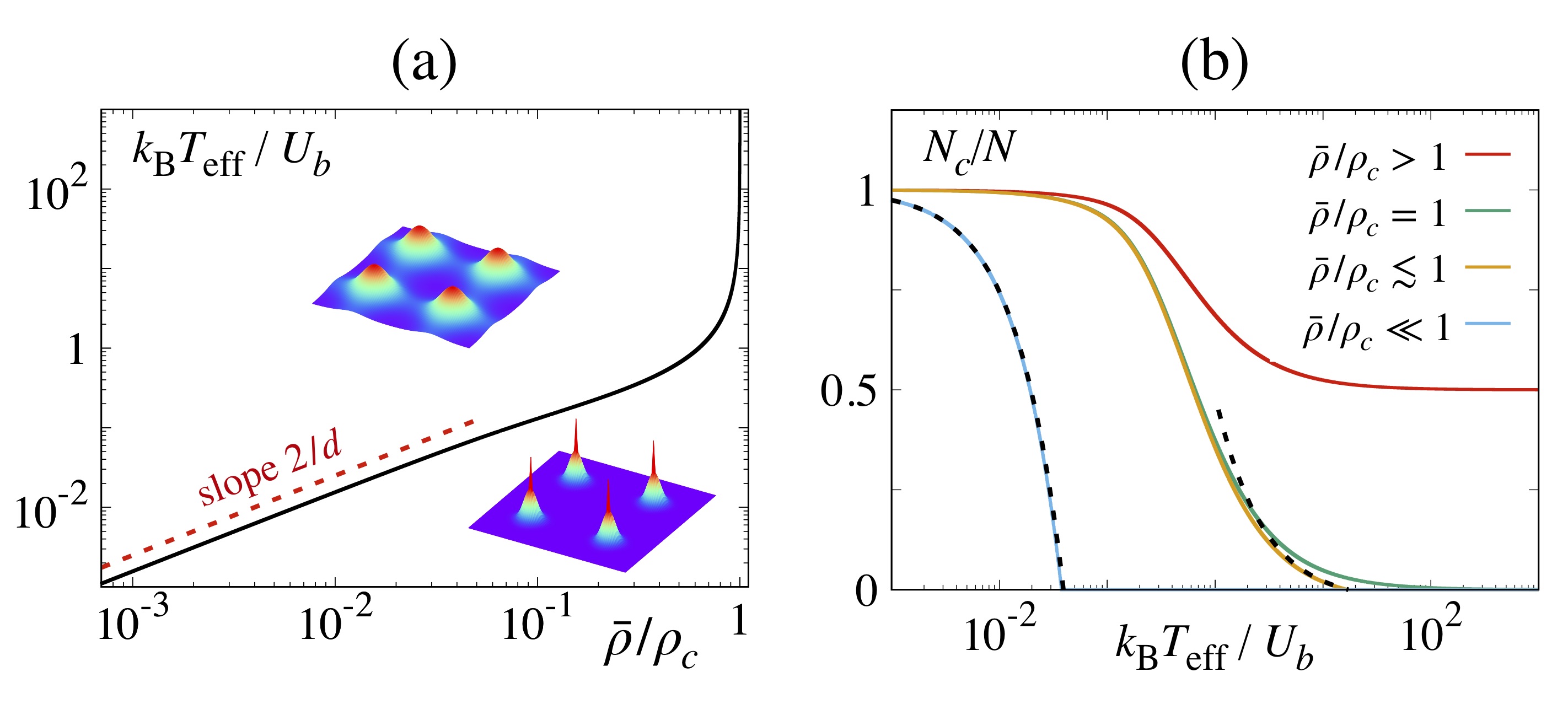}
	\caption{Transition to condensation in $d=2$; no qualitative differences are expected in other dimensions.
	(a) Phase diagram of the system in the reduced mean density $\bar{\rho}/\rho_c$ and effective temperature
	$k_{\rm B} T_{\rm eff}/U_b$ plane. 
	The continuous black line marks the transition corresponding to $T_{\rm eff} = T_c$ defined from \Eref{eq_N_cos} by setting
	$N_c/N = 0$.
	(b) Condensate fraction as a function of the reduced effective temperature for several values of the ratio $\bar{\rho}/\rho_c$.
	The dashed black lines indicate the approximate behaviour of $N_c/N$ for $k_{\rm B} T_{\rm eff} \ll U_b$ (\Eref{eq_cond_frac_cos_high}) and
	$k_{\rm B} T_{\rm eff} \gg U_b$ (\Eref{eq_cond_frac_cos_low}), respectively.
	 }
	\label{fig_transition_cos}
\end{figure}


When $\rho_0$ becomes larger than the diffusivity edge, some of the particles form a condensate in the ground state. In this case, the normalization of the density profile is given by the second line of \Eref{eq_N_cos}. In the high energy barrier limit, the condensate fraction can be approximated by 
\begin{equation}
\frac{N_c}{N} \underset{k_{\rm B}T_{\rm eff} \ll U_b}{\sim} 1 - \left(\frac{T_{\rm eff}}{T^0_c}\right)^{\frac{d}{2}} \,,
\label{eq_cond_frac_cos_high}
\end{equation}
where the effective transition temperature reads $T^0_c = \frac{k}{2\pi k_{\rm B}} \left(n / \rho_c \right)^{\frac{2}{d}}$ \cite{Golestanian2019BEC}.
In this limit, $N_c/N$ takes a similar form as in the case of a free ideal Bose gas \cite{Ziff1977BoseGas}. Defining $\bar{\rho} \equiv N/L^d$ as the average density of particles, the condensate fraction in the shallow potential limit reads
\begin{equation}
\frac{N_c}{N} \underset{k_{\rm B}T_{\rm eff} \gg U_b}{\sim} \left(1 - \frac{\rho_c}{\bar{\rho}} \right)\left(1 - \frac{T^\infty_c}{T_{\rm eff}}\right) \,,
\label{eq_cond_frac_cos_low}
\end{equation}
where $T^\infty_c = U_b\left[2k_{\rm B}\left(1- \bar{\rho}/\rho_c\right)\right]^{-1}$. The fact that the transition temperature diverges when $\bar{\rho}$ approaches $\rho_c$ is due to the finiteness of the barrier height $U_b$, which leads to flat density profiles at high effective temperatures. As shown in the phase diagram of Figure ~\ref{fig_transition_cos}(a), the phase behaviour for $k_{\rm B}T_{\rm eff} / U_b \gg 1$ is only set by the ratio $\bar{\rho} / \rho_c$.

The scaling of the condensate fraction as a function of $k_{\rm B}T_{\rm eff} / U_b$ is shown in Figure~\ref{fig_transition_cos}(b). If $\bar{\rho} > \rho_c$, the transition is suppressed, and $N_c/N$ goes from $1$ at vanishing $T_{\rm eff}$ to finite values when $k_{\rm B}T_{\rm eff} / U_b \to \infty$.
When $\bar{\rho} < \rho_c$, there exists a finite effective temperature $T_c$ for which $N_c/N$ reaches $0$, and above which no condensation occurs. If $\bar{\rho} \ll \rho_c$, the transition happens at small effective temperatures and is similar to BEC. On the other hand, for $\rho_c \gtrsim \bar{\rho}$ the exponent associated to the scaling of the condensate fraction is equal to $-1$ in any dimension (see \Eref{eq_cond_frac_cos_low}). Finally, in the particular case of $\bar{\rho} = \rho_c$, we find that $N_c/N \sim (k_{\rm B}T_{\rm eff} / U_b)^{-1}$ at large $k_{\rm B}T_{\rm eff} / U_b$, such that the transition temperature is located exactly at infinity. 

Although deriving an analogue to \Eref{eq_N_cos} for arbitrary functions $D(\rho)/M(\rho)$ is out of the scope of this work, \ref{app_Cond_frac} shows how some progress can be achieved in the limit $k_{\rm B}T_{\rm eff} \gg U_b$. Indeed, assuming that the diffusivity edge is approached as 
\begin{equation}
D(\rho)/M(\rho) \underset{\rho \to \rho_c}{\sim} (1 - \rho / \rho_c)^{z-1} \,,
\label{eq_def_z}
\end{equation}
where $z \ge 1$, the associated scaling of the condensate fraction reads 
\begin{equation}
\frac{N_c}{N} \underset{k_{\rm B}T_{\rm eff} \gg U_b}{\sim} \left(1 - \frac{\rho_c}{\bar{\rho}}\right) \left[ 1- \left(\frac{T^\infty_c}{T_{\rm eff}}\right)^{\frac{1}{z}}  \right] \,,
\label{eq_cond_frac_cos_low_z}
\end{equation}
with $k_{\rm B}T^\infty_{c} / U_b \sim (1-\bar{\rho}/\rho_c)^{-z}$.
For shallow potentials the condensate fraction exponent therefore takes a nonuniversal value, which is set by how fast the diffusivity edge is reached. This result is in clear departure from \Eref{eq_cond_frac_cos_high} for $k_{\rm B}T_{\rm eff} \ll U_b$, where the exponent $\frac{d}{2}$ remains independent of the shape of $D(\rho)/M(\rho)$ \cite{Golestanian2019BEC}.

\section{Generalized thermodynamics}

\label{sec_Thermodynamics_cos}

\begin{figure}[t!]
	\centering
	\includegraphics[width = \textwidth]{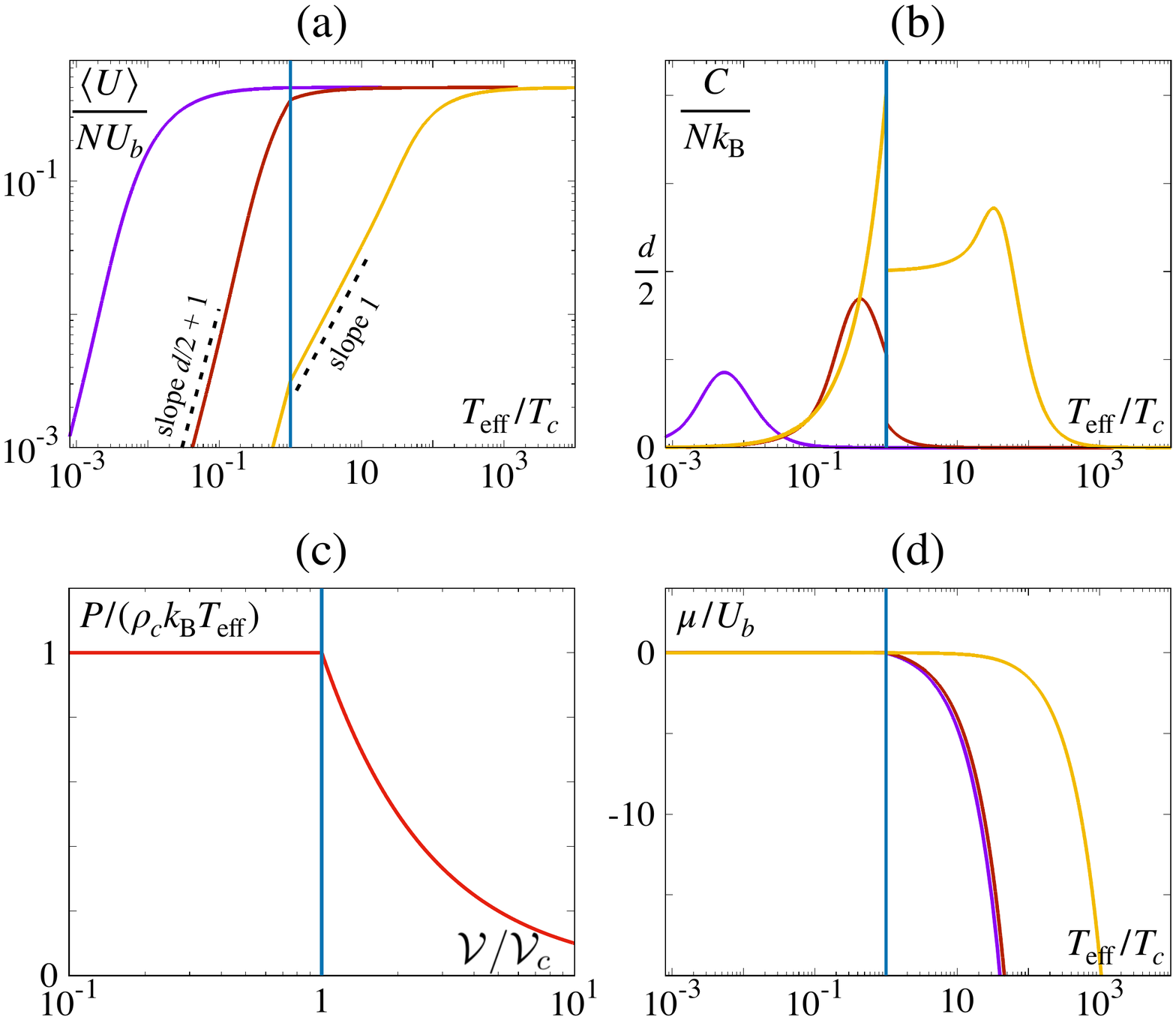}
	\caption{Thermodynamics of the system in $d=2$; no qualitative differences are expected in other dimensions.
	(a,b) Mean potential energy $\langle U \rangle$ and heat capacity $C$ as functions of $T_{\rm eff}/T_c$ for 
	$\bar{\rho} / \rho_c = 0.99$ (purple), 0.5 (red), and 0.002 (yellow).
	(c) Typical isotherm of the pressure showing a plateau for effective volumes ${\cal V} \le  {\cal V}_c$.
	(d) Chemical potential as a function of $T_{\rm eff}/T_c$; the different lines correspond to the same cases as for (a,b).
	In all panels the vertical blue line locates the transition.
	 }
	\label{fig_thermo_cos}
\end{figure}

We now turn to the construction of a generalized thermodynamic formalism for the system. The average potential energy $\langle U \rangle \equiv \int \rmd^d {\bm r}\, U({\bm r}) \rho\left(U({\bm r})\right)$ reads 
  \begin{equation}
 \langle U \rangle = 
 \label{eq_U_cos}
 \cases{ \frac{N U_b}{2} \left (1 - \frac{I_1\left( \frac{\beta U_b}{2d}\right)}{I_0\left( \frac{\beta U_b}{2d}\right)} \right) 
 & $T_{\rm eff} > T_c$ \\
\frac{\left(N - N_c\right)U_b}{2} \left (1 - \frac{I_1\left( \frac{\beta U_b}{2d}\right)}{I_0\left( \frac{\beta U_b}{2d}\right)} \right)
& $T_{\rm eff} \le T_c$} \,.
 \end{equation}
A heat capacity can then be defined from the mean energy via $C \equiv \rmd\langle U \rangle / \rmd T_{\rm eff}$. For the present system, we find the following expressions after some algebra
   \begin{equation}
\fl C = \cases{\frac{N k_{\rm B}}{4d} (\beta U_b)^2 \left[ 1 - \frac{I_1\left( \frac{\beta U_b}{2d}\right)}{I_0\left( \frac{\beta U_b}{2d}\right)} 
\left(  \frac{2d}{\beta U_b} + \frac{I_1\left( \frac{\beta U_b}{2d}\right)}{I_0\left( \frac{\beta U_b}{2d}\right)} \right) \right]
 & $T_{\rm eff} > T_c$ \\
 \frac{\left(N - N_c\right)k_{\rm B}}{4d} (\beta U_b)^2
 & \nonumber \\
\times \left\{ d + 1 - \frac{I_1\left( \frac{\beta U_b}{2d}\right)}{I_0\left( \frac{\beta U_b}{2d}\right)}  
  \left[  2d \left( 1 + \frac{1}{\beta U_b} \right) - (d-1) \frac{I_1\left( \frac{\beta U_b}{2d}\right)}{I_0\left( \frac{\beta U_b}{2d}\right)} \right] \right\}
 & $T_{\rm eff} \le T_c$} \,.
 \end{equation}
The change in the heat capacity at the transition, $\Delta C \equiv C(T = T_c^-) - C(T = T_c^+)$ is then given by
\begin{equation}
\Delta C = \frac{N k_{\rm B}}{4} (\beta_c U_b)^2 \left[ 1 + \frac{I_1\left( \frac{\beta_c U_b}{2d}\right)}{I_0\left( \frac{\beta_c U_b}{2d}\right)} 
\left( \frac{I_1\left( \frac{\beta_c U_b}{2d}\right)}{I_0\left( \frac{\beta_c U_b}{2d}\right)} - 2 \right) \right] \,,
\end{equation}
with $\beta_c \equiv (k_{\rm B}T_c)^{-1}$. Generally, $\Delta C$ is nonzero such that the heat capacity experiences a discontinuous jump at the transition (see Figure~\ref{fig_thermo_cos}(b)). For BEC in free space, this feature appears only for $d \ge 5$ \cite{Ziff1977BoseGas}, while it can be affected by confinement \cite{Bagnato1987BECPotential,Dalfovo1999RMP}. Similar features are expected for the diffusivity edge problem, where the shape of $D(\rho)/M(\rho)$ in the vicinity of $\rho_c$ could additionally play a role. These questions will be addressed in a separate publication \cite{FollowPaper}.
 

Using the asymptotic expansions of the modified Bessel functions \eref{eq_Bessel}, the analytical expressions for $\langle U \rangle$ and $C$ can be obtained in the strong and weak confinement limits. As shown in Figures~\ref{fig_thermo_cos}(a,b), for $k_{\rm B}T_{\rm eff} \ll U_b$ their behaviour corresponds to that of an ideal Bose gas \cite{Golestanian2019BEC}, while for $k_{\rm B}T_{\rm eff} \gg U_b$, we obtain
\begin{eqnarray}
\fl \langle U \rangle \underset{k_{\rm B}T_{\rm eff} \gg U_b}{\sim} \frac{N U_b}{2} \,, 
\qquad & C \underset{k_{\rm B}T_{\rm eff} \gg U_b}{\sim} \frac{N k_{\rm B}}{8d} (\beta U_b)^{2} 
& T_{\rm eff} > T_c\,, \\
\fl \langle U \rangle \underset{k_{\rm B}T_{\rm eff} \gg U_b}{\sim} \frac{\rho_c}{\bar{\rho}}\frac{N U_b}{2} \,, 
\qquad & C \underset{k_{\rm B}T_{\rm eff} \gg U_b}{\sim} \left(d + \frac{1}{2}\right)\frac{\rho_c}{\bar{\rho}}\frac{N k_{\rm B}}{4d}(\beta U_b)^{2} \qquad 
&T_{\rm eff} \le T_c\,.
\end{eqnarray} 
In the limit of a shallow potential and a high effective temperature, $\langle U \rangle$ becomes independent of $T_{\rm eff}$ and scales linearly with $U_b$. The heat capacity thus vanishes as $(\beta U_b)^{2}$. Note that the functions below $T_c$ are proportional to the ratio $\rho_c/\bar{\rho}$, which highlights the fact that particles in the condensate do not contribute to the total energy.
  
A thermodynamic entropy can be defined for the system as $\rmd S \equiv \rmd\langle U \rangle/T_{\rm eff}$. After some algebra, we find the following expressions
\begin{equation}
\label{eq_S_cos}
\!\!\!\!\!\!\!\! \!\!\!\!  S = \cases{\frac{N k_{\rm B}}{2} \left\{ 2 \left[ 1 - \ln\left(\frac{\rho_0}{\rho_c}\right)\right] 
+ \beta U_b \left[ 1 - \frac{I_1\left(\frac{\beta U_b}{2d}\right)}{I_0\left(\frac{\beta U_b}{2d}\right)} \right] \right\}
 & $T_{\rm eff} > T_c$ \\
\frac{\left(N - N_c\right)k_B}{2}\left[ 2 + \beta U_b \left(1 - \frac{I_1\left(\frac{\beta U_b}{2d}\right)}{I_0\left(\frac{\beta U_b}{2d}\right)} \right) \right]
 & $T_{\rm eff} \le T_c$} \,.
\end{equation}
We note that the same result can be derived from a Gibbs definition of generalized entropy, which is consistent with \Eref{eq_solU} as the relationship between the energy and the probability measure. A similar definition has been introduced in Ref.~\cite{Chavanis2008FP}. In case with a large energy barrier, we find that the entropy exhibits an ideal Bose gas behaviour \cite{Golestanian2019BEC}.
On the other hand, for $k_{\rm B}T_{\rm eff} \gg U_b$ the energy states are distributed uniformly in space, and $S/L^d\simeq -k_{\rm B}\rho[\ln(\rho/\rho_c) - 1]$, with distributions $\rho = \bar{\rho}$ $(T_{\rm eff} > T_c)$ and $\rho = \rho_c$ $(T_{\rm eff} \le T_c)$.

A remarkable feature of BEC concerns the divergence of the isothermal compressibility at the transition. A thermodynamic pressure can be defined for the system from a generalized Helmholtz free energy ${\cal F} \equiv \langle U \rangle - T_{\rm eff} S$. The typical volume ${\cal V}$ of the confined system can be obtained from dimensional analysis: ${\cal V} \equiv N/\rho_0 = (N-N_c)/\rho_c$. Using \Eref{eq_N_cos}, this reads\footnote{An estimate of the ``volume'' $\ell^d$ occupied by the particles can also be obtained from $\ell^2 \sim \frac{\int \rmd^d {\bm r}\, |{\bm r}|^2 \exp[-\beta U({\bm r})]}{\int \rmd^d {\bm r}\, \exp[-\beta U({\bm r})]}$. While this integral cannot be solved analytically, its asymptotic forms in the shallow and deep potential limits correspond to those of \Eref{eq_Vol_cos} up to constant pre-factors.}
\begin{equation}
\label{eq_Vol_cos}
{\cal V} = L^d \exp\left(-\frac{\beta U_b}{2} \right) I_0^d\left(\frac{\beta U_b}{2d} \right) \,.
\end{equation}
When $k_{\rm B}T_{\rm eff} \ll U_b$, we find ${\cal V} \sim \lambda^d = (2 \pi k_{\rm B} T_{\rm eff} / k)^{\frac{d}{2}}$, which corresponds to the typical volume occupied by an ideal Bose Gas confined in a harmonic potential \cite{Romero2005HP}, while for $k_{\rm B}T_{\rm eff} \gg U_b$, it is simply given by the size of the system $L^d$.

Defining $P_{\rm therm}$ as the conjugate variable to ${\cal V}$ provides the following equations of state
\begin{equation}
\label{Pth_cos}
P_{\rm therm} \equiv - \left. \left( \frac{\partial {\cal F}}{\partial {\cal V}} \right) \right|_{T_{\rm eff},N} = \cases{ \frac{N k_{\rm B}T_{\rm eff}}{\cal V}  & $T_{\rm eff} > T_c$ \\ 
\rho_c k_{\rm B}T_{\rm eff} & $T_{\rm eff} \le T_c$}\,.
\end{equation} 

Equations \eref{Pth_cos} are identical to those derived in~\cite{Golestanian2019BEC}, and stress again the similarities with BEC. 
These expressions can moreover be derived from a mechanical definition of the pressure, which we denote as $P_{\rm mech}$. 
Thanks to the periodicity of the potential $U$, the forces that it induces do not create pressure difference between adjacent unit cells, 
such that a bulk can be defined for this system.
Following previous works \cite{Solon2015Pressure}, 
the bulk mechanical pressure is defined as the average force per unit surface area 
exerted by the particles on a potential $W$ which confines the system in a finite volume $L^d$.
From the symmetries of the problem, the calculation is moreover carried out in one dimension.
Assuming that the edge of the system corresponds to a dip of $U$\footnote{This choice, made for convenience, 
does not affect the result of the calculation as long as the system remains made of an integer number of unit cells.}, 
$W(r)$ is monotonously growing with $r$, satisfies $W(r) = 0$ for $r \le L$ and $W(r \to \infty) \to \infty$ 
(it is therefore assumed that $U = 0$ outside the sample).
This way, $P_{\rm mech}$ reads
\begin{equation}
P_{\rm mech} = \int_{L}^{\infty} \rmd r \, \rho\left(W(r)\right) \partial_{r} W(r) = \int_{0}^{\infty} \rmd W \, \rho\left(W\right) \,,
\end{equation}
which after replacing $\rho$ by its expression as function of the potential \eref{eq_Bweights_rho}, 
leads to $P_{\rm mech} = P_{\rm therm} \equiv P$.
As at equilibrium, and in a limited number of nonequilibrium cases \cite{Solon2015Pressure}, this result is moreover independent of the details of $W$. 


Let us consider a typical isotherm of $P$ as shown in Figure~\ref{fig_thermo_cos}(c). Defining ${\cal V}_c \equiv N / \rho_c$ as the volume below which condensation occurs, we find that for ${\cal V} > {\cal V}_c$ the system possesses an ideal gas equation of state and $P$ scales like ${\cal V}^{-1}$.
For ${\cal V} \le {\cal V}_c$ the pressure becomes independent of ${\cal V}$ and the corresponding isotherm exhibits a plateau, such that the isothermal compressibility of the system, $\kappa_{T_{\rm eff}} = - {\cal V}^{-1} \left( \frac{\partial P}{\partial {\cal V}} \right)^{-1}_{T_{\rm eff}}$, diverges
at the threshold.

We end this section by computing the generalized chemical potential $\mu$, defined as the conjugate variable to $N$. From Equations \eref{eq_U_cos} and \eref{eq_S_cos}, we find
\begin{equation}
\label{mu_cos}
\mu \equiv \left.\left( \frac{\partial {\cal F}}{\partial N} \right) \right|_{T_{\rm eff},{\cal V}} = \cases{ k_{\rm B}T_{\rm eff} \ln\left(\frac{\rho_0}{\rho_c}\right) & $T_{\rm eff} > T_c$ \\ 
0 & $T_{\rm eff} \le T_c$}\,.
\end{equation} 
We thus find that $\mu$ vanishes at the transition where $\rho_0 = \rho_c$ and remains identically $0$ in the condensate phase, as shown in Figure \ref{fig_thermo_cos}(d). From Equations \eref{eq_Bweights_rho} and \eref{mu_cos}, the density profile outside the ground state thus takes the general form
 \begin{equation}
\rho(U > 0) = \rho_c \exp\left[\beta (\mu - U)\right] \,,
\end{equation} 
above and below the transition.


\section{Concluding remarks} \label{sec_discussion}

We have studied the consequences of having a diffusivity edge for a system of particles embedded in a sinusoidal potential in arbitrary dimensions. This configuration leads to the formation multiple coexisting condensates. We have identified two asymptotic regimes that exhibit qualitatively different properties. At low effective temperatures or high potential barriers, the behaviour of the system is analogous to that of an ideal Bose gas in free space, similarly to the case treated in Ref. \cite{Golestanian2019BEC} considering a single harmonic trap. For shallow potentials, qualitative features such as the presence of a transition, at which the heat capacity is discontinuous, as well as the divergence of the isothermal compressibility and the vanishing of the chemical potential in the condensed phase, persist when the mean density stays lower than the threshold $\rho_c$. 

We have, however, uncovered quantitative differences is this case. For example, we have found that for $D(\rho)/M(\rho) \sim (\rho_c - \rho)^{z-1}$ near $\rho_c$ the scaling of the condensate fraction with the effective temperature takes an exponent of $-z^{-1}$ (see \Eref{eq_cond_frac_cos_low_z}). The exponent $z$ is also expected to affect the scaling of other functions. For instance, using the results derived in \ref{app_Cond_frac}, it is possible to show that $C {\sim} (\beta U_b)^{1 + z^{-1}}$ as $\beta U_b \to 0$. A systematic study of the effect of $z$ and of the shape of the potential is relegated to future publications \cite{FollowPaper}.

The derivation of our results relies on the hypothesis that particles are divided evenly among cells of volume $(2 r_b)^d$. However, a vanishing diffusion coefficient may seem ``pathological'' at the mean field level considered here, as the absence of fluctuations would lead to a breakdown of ergodicity. It should thus be stressed that the range of validity of the results derived here concerns all systems described by \Eref{eq_continuity} and for which the density-dependent hydrodynamic coefficients result from the integration of various microscopic processes, e.g. interactions, while fluctuations, although possibly weak, remain present.

\begin{appendix}


\section{Derivation of the condensate fraction for $\beta U_b \to 0$ and arbitrary diffusion}

\label{app_Cond_frac}

This section is devoted to the derivation of \Eref{eq_cond_frac_cos_low_z} describing the weak confinement 
behaviour of the condensate fraction assuming a general functional form of $D(\rho)/M(\rho)$ near $\rho_c$.
Below $T_c$, the potential can be formally written as $\beta U(\rho) \equiv u(\rho/\rho_c)$, 
where this rescaled form satisfies
\begin{equation}
u(s) = -\int_1^s \rmd t \, \frac{y(t)}{t}
\label{app_eq_u}
\end{equation} 
with $y(\rho/\rho_c) \equiv \beta D(\rho)/M(\rho)$.
Denoting $\rho_b \equiv \rho(U_b)$,
from \Eref{app_eq_u} the limit $u(\rho_b / \rho_c) = \beta U_b \to 0$ is attained for $\rho_b \to \rho_c$.
In the following, we assume that the diffusivity edge is reached at $\rho = \rho_c$
following a power law with an exponent $z - 1 \ge 0$:
\begin{equation}
y(s) \underset{s \to 1^-}{\sim} y_0 (1-s)^{z-1} \,,
\label{eq_y}
\end{equation} 
where $y_0$ is a constant and $y(s) = 0$ for all $s \ge 1$.
The Taylor expansion of $u(s)$ in $s=1$ reads
\begin{equation}
u(s) = \sum_{n=0}^{\infty} \frac{u^{(n+1)}(1)}{(n+1)!} (s-1)^{n+1} \,,
\label{eq_Taylor_def_u}
\end{equation}
where $u^{(n+1)}$ stands for the $(n+1)^{\rm th}$ derivative of $u$, 
and is obtained from \Eref{app_eq_u} as
\begin{eqnarray}
u^{(n+1)}(s) & = & -\sum_{p=0}^{n} {n \choose p} y^{(p)}(s) \left(s^{-1}\right)^{(n-p)} \,, \\
u^{(n+1)}(s) &  \underset{s \to 1^-}{\sim} & y_0 
\sum_{p=0}^{{\rm min}(z-1,n)}  {z-1 \choose p} n! (-1)^{n+1} (1-s)^{z- 1 - p} \,.
\label{eq_unp1}
\end{eqnarray}
It is clear from \Eref{eq_unp1} that $u^{(n+1)}(s)$ will cancel when $s \to 1^-$ for all $n < z-1$,
and that $u^{(z + i)}(1) = y_0 (z + i - 1)!(-1)^{z + i}$ for all $i \ge 0$.
Inserting this expression in \Eref{eq_Taylor_def_u}, we get
\begin{equation}
u(s) = y_0 \sum_{n=0}^{\infty} \frac{(1-s)^{z+n}}{z+n} \,,
\label{eq_Taylor_u}
\end{equation}
which corresponds as expected to $u(s) = -y_0 \ln(s)$ for $z=1$. 
We then get at leading order for $s \lesssim 1$
\begin{equation}
u(s) \underset{s \to 1^-}{\sim} y_0 \frac{(1-s)^{z}}{z} + \Or\left((1-s)^{z+1}\right) \,.
\label{eq_u_first_order}
\end{equation}
Therefore, inverting this expression and coming back to the initial variables we find outside the ground state
\begin{equation}
\rho(U > 0) \underset{\beta U_b \to 0}{\sim} \rho_c \left[ 1 - \left( \frac{\beta U z}{y_0}\right)^{\frac{1}{z}} \right] 
\label{eq_rho_U}
\end{equation}

Using \Eref{eq_rho_U} and the definition of the potential 
$U({\bm r}) = \frac{U_b}{2} \left[1 - d^{-1} \sum_k \cos(\pi r_k / r_b) \right]$, the normalization in the condensation phase thus obeys
\begin{equation}
N = N_c + \rho_c L^d \left[ 1 -  \left( \frac{\beta U_b z}{2y_0}\right)^{\frac{1}{z}} {\cal G}(z) \right] \,,
\label{app_normalization}
\end{equation}
where the function ${\cal G}(z) = \int_0^1 \rmd x_1 \ldots \int_0^1 \rmd x_d\, [1 - d^{-1}\sum_k \cos(\pi x_k)]^{1/z}$
is nonzero and analytic but has no simple expression in general.
Defining $k_{\rm B}T_c = z U_b/ ( 2 y_0) {\cal G}^{z}(z) (1-\bar{\rho}/\rho_c)^{-z}$, \Eref{app_normalization}
is finally recast as \Eref{eq_cond_frac_cos_low_z}.

\end{appendix}

\section*{References}

\bibliographystyle{iopart-num}
\bibliography{Biblio}

\end{document}